\documentstyle[12pt,psfig]{article}

\begin{document}

\newcommand{\ep}{equivalence principle~}
\newcommand{\Ch}{Chandrasekhar~}
\newcommand{\Chp}{Chandrasekhar}
\newcommand{\Sc}{Schwarzschild~}
\newcommand{\Scp}{Schwarzschild}
\newcommand{\Sw}{Schwarzschild~}
\newcommand{\Swp}{Schwarzschild}
\newcommand{\Sch}{Schr{\"{o}}dinger~}
\newcommand{\Schp}{Schr{\"{o}}dinger}
\newcommand{\OVp}{Oppenheimer--Volkoff}
\newcommand{\OV}{Oppenheimer--Volkoff~}
\newcommand{\GR}{General Relativity~}
\newcommand{\GT}{General Theory of Relativity~}
\newcommand{\GRp}{General Relativity}
\newcommand{\GTp}{General Theory of Relativity}
\newcommand{\STp}{Special Theory of Relativity}
\newcommand{\ST}{Special Theory of Relativity~}
\newcommand{\Lt}{Lorentz transformation~}
\newcommand{\Ltp}{Lorentz transformation}
\newcommand{\rel}{relativistic~}
\newcommand{\relp}{relativistic}
\newcommand{\msun}{M_{\odot}}
\newcommand{\eos}{equation of state~}
\newcommand{\eoss}{equations of state~}
\newcommand{\eossp}{equations of state}
\newcommand{\eosp}{equation of state}
\newcommand{\Eos}{Equation of state}
\newcommand{\Eosp}{Equation of state}
\newcommand{\beqn}{\begin{eqnarray}}
\newcommand{\eeqn}{\end{eqnarray}}
\newcommand{\nonum}{\nonumber \\}
\newcommand{\walecka}{$\sigma,\omega,\rho$~}
\newcommand{\waleckap}{$\sigma,\omega,\rho$}
\newcommand{\bbar}[1] {\mbox{$\overline{#1}$}} 
\newcommand{\courtesy}{~Reprinted with permission of Springer--Verlag 
New York; copyright 1997}
\newcommand{\oo}{{\"{o}}}
\newcommand{\au}{{\"{a}}}


\newcommand{\approxlt} {\mbox {$\stackrel{{\textstyle<}} {_\sim}$}}
\newcommand{\approxgt} {\mbox {$\stackrel{{\textstyle>}} {_\sim}$}}
\newcommand{\mearth} {\mbox {$M_\oplus$}}
\newcommand{\rearth} {\mbox {$R_\oplus$}}
\newcommand{\eo} {\mbox{$\epsilon_0$}}
\newcommand{\bfalpha} {\mbox{\mbox{\boldmath$\alpha$}}}
\newcommand{\bfgamma} {\mbox{\mbox{\boldmath$\gamma$}}}
\newcommand{\bfrho} {\mbox{\mbox{\boldmath$\rho$}}}
\newcommand{\bfsigma} {\mbox{\mbox{\boldmath$\sigma$}}}
\newcommand{\bftau} {\mbox{\mbox{\boldmath$\tau$}}}
\newcommand{\bfLambda} {\mbox{\mbox{\boldmath$\Lambda$}}}
\newcommand{\bfpi} {\mbox{\mbox{\boldmath$\pi$}}}
\newcommand{\bfomega} {\mbox{\mbox{\boldmath$\omega$}}}
\newcommand{\bp} {\mbox{\mbox{\boldmath$p$}}}
\newcommand{\br} {\mbox{\mbox{\boldmath$r$}}}
\newcommand{\bx} {\mbox{\mbox{\boldmath$x$}}}
\newcommand{\bv} {\mbox{\mbox{\boldmath$v$}}}
\newcommand{\bu} {\mbox{\mbox{\boldmath$u$}}}
\newcommand{\bk} {\mbox{\mbox{\boldmath$k$}}}
\newcommand{\bA} {\mbox{\mbox{\boldmath$A$}}}
\newcommand{\bB} {\mbox{\mbox{\boldmath$B$}}}
\newcommand{\bF} {\mbox{\mbox{\boldmath$F$}}}
\newcommand{\bI} {\mbox{\mbox{\boldmath$I$}}}
\newcommand{\bJ} {\mbox{\mbox{\boldmath$J$}}}
\newcommand{\bK} {\mbox{\mbox{\boldmath$K$}}}
\newcommand{\bP} {\mbox{\mbox{\boldmath$P$}}}
\newcommand{\bS} {\mbox{\mbox{\boldmath$S$}}}
\newcommand{\bdel} {\mbox{\mbox{\boldmath$\bigtriangledown$}}}

\newcommand{\eps} {\mbox {$\epsilon$}}
\newcommand{\gpercm} {\mbox {${\rm g}/{\rm cm}^{3}$}}
\newcommand{\rhon} {\mbox {$\rho_{0}$}}
\newcommand{\rhoc} {\mbox {$\rho_{c}$}}
\newcommand{\fmm} {\mbox {${\rm fm}^{-3}$}}
\newcommand{\bag} {\mbox {$B^{1/4}$}}

\newcommand{\fraca} {\mbox {$ \frac{1}{2}       $}}
\newcommand{\fracb} {\mbox {$ \frac{3}{2}       $}}
\newcommand{\fracc} {\mbox {$ \frac{1}{4}       $}}
\newcommand{\fraccc} {\mbox {$ \frac{1}{3}       $}}

\newcommand{\tit}
{Neutron Star Constraints on the H Dibaryon}

\newcommand{\auth} {Norman K. Glendenning and J\"urgen Schaffner-Bielich}
\newcommand{\lbl}{LBL-41396}
\newcommand{\dateofdoc}{\today}
\newcommand{\adr} 
{Nuclear Science Division \& 
Institute for Nuclear and Particle Astrophysics,
  Lawrence Berkeley  National Laboratory,
   MS: 70A-3307 \\ Berkeley, California 94720}

\newcommand{\doe}
{This work was supported by the
Director, Office of Energy Research,
Office of High Energy
and Nuclear Physics,
Division of Nuclear Physics,
of the U.S. Department of Energy under Contract
DE-AC03-76SF00098.}

\newcommand{\ect}{A part of this work was done at the ECT*,
Villa Tambosi, Trento, Italy.}


\begin{titlepage}
\begin{center}
\parbox{3in}{\begin{flushleft}Preprint \end{flushleft}}%
\parbox{3in}{\begin{flushright} \lbl  \end{flushright}}
~\\[7ex]

\begin{Large}
\tit {\footnote{\doe}}\\[2ex]
\end{Large}
\renewcommand{\thefootnote}{\fnsymbol{footnote}}
\setcounter{footnote}{1}
~~\footnotetext{\tiny{[nkg.papers]sample.tex,  \today} }

\begin{large}
\auth\\[3ex]
\end{large}
\adr\\[3ex]
\dateofdoc \\[3ex]
\end{center}
\begin{figure}[htb]
\vspace{-.8in}
\begin{center}
\leavevmode
\hspace{-.2in}
\psfig{figure=ps.comp_h_k240m78_su6_uhm30,%
width=2.66in,height=3.2in}
\end{center}
\end{figure}

\begin{center} {\bf PACS} \\ 
26.60.+c,~~14.20.Pt,~~98.38.Mz,~~97.60.Jd,~~97.60.Gb 
\end{center}
\end{titlepage}

\clearpage


\begin{center}
\begin{Large}
\tit \\[7ex]
\end{Large}

\begin{large}
\auth \\[2ex]
\end{large}
\adr
\end{center}


\begin{abstract}
We study the influence of a possible H
dibaryon condensate on the equation of state and the overall
properties of neutron stars whose population otherwise contains
nucleons and hyperons.
In particular, we are interested in the question of whether neutron stars
and their masses can be used to say anything about the existence and
properties of the H dibaryon.
We find that the equation of state is softened by the
appearance of a dibaryon condensate and can result in
a mass plateau for neutron stars. If the limiting neutron star mass
is about that of the Hulse-Taylor pulsar
a  condensate of H dibaryons of vacuum mass $\sim 2.2$ GeV
and a moderately attractive potential in the medium
could not be ruled out.
On the other hand, if the medium potential were even moderately repulsive,
the H,  would not likely exist  in neutron stars.
If neutron stars of mass $\sim 1.6 M_\odot$ were known to exist,
attractive medium effects  for the H could be ruled out.
\end{abstract}

\section{Introduction}

Since Jaffe proposed that there  may exist a stable dihyperon
(a quark composite with baryon number two)
\cite{Jaffe77}, an ongoing  quest for this particle began
\cite{Carroll78}. Recent searches using  
kaon beams \cite{Aoki90} or heavy ion beams \cite{Belz96,Belz97,Stotz97} 
found no candidates or are still in progress \cite{Craw98}. 
There exist some claims for evidence for the H
dibaryon produced in proton-nucleus \cite{Shahba93} and in heavy-ion collisions
\cite{Long95}. Nevertheless, these candidates might be misidentified $K^0_L$
as seen in \cite{Belz97}. For a most recent overview on the search for 
the H dibaryon we refer to \cite{HYP97}.

There are numerous mass estimates for the H dibaryon 
and they are reviewed in \cite{Dover89}. 
The existence or nonexistence of the H
dibaryon is strongly 
connected with the observation of double $\Lambda$ hypernuclei which has been
discussed in \cite{Dalitz89}. 
Three double
$\Lambda$ hypernuclei have been reported in literature:
$^{~~6}_{\Lambda\Lambda}$He \cite{Danysz63},
$^{~10}_{\Lambda\Lambda}$Be \cite{Prowse66}, and
$^{~13}_{\Lambda\Lambda}$B \cite{Aoki91,Dover91}.
The two $\Lambda$'s can decay by strong interactions to the H dibaryon.
As this has not been seen in the above events, the H must either be heavier than
$m_H> 2 m_\Lambda+B_{\Lambda\Lambda}\approx 2.22$ GeV \cite{Kerb84} 
or the events are
misidentified as an H hypernucleus with a shallow attractive nuclear potential
\cite{Dover89}. 
A more stringent condition is the observation of the weak mesonic decay of the
double $\Lambda$ hypernuclei giving $m_H> m_\Lambda + m_p + m_{\pi^-} +
B_\Lambda \approx 2190$ MeV \cite{Jaffe91} where $B_\Lambda$ depends on the
mass of the decay fragment and is $B_\Lambda=-3.1$ MeV for $^5_\Lambda$He.
In all cases, a deeply bound H dibaryon seems to be ruled out by these events. 

If the H dibaryon exists, it will have a
 certain impact also on the properties of
dense matter. It is quite established nowadays, that neutron stars have a large
hyperon fraction in the core and  might be described as giant
hypernuclei, though bound by gravity
 \cite{Glen85}. Here again, the presence of hyperons might restrict
certain properties of the H dibaryon.
Recently, studies for neutron stars have been done for nuclear matter without
hyperons but including H dibaryon condensation
\cite{Fae97a} and 
limits have been set for the coupling constants of the H
dibaryon \cite{Fae97b}.

There might exist heavier partners of the H dibaryon, lumps of strange quark
matter dubbed strangelets. There are several heavy-ion experiments dedicated 
to search for this novel form of matter \cite{Arm97,Beavies95,Apple96}.
In the MIT bag model, strangelets with $A\leq 6$ are
found to be unbound \cite{Aerts78}. Nevertheless, light 
strangelet candidates in the range of $6<A<40$  
might be stable against weak hadronic decay \cite{Gilson93,Scha97}
(an overview of the properties of strange matter can be found in \cite{GS96}).
The H dibaryon as well as these light strangelets can occur in dense matter as
a precursor of the phase transition to a quark plasma. 

In this paper, we study the influence of H dibaryons and other strangelet
candidates on the composition and structure of neutron stars including the
hyperon degree of freedom.
We are particularly interested in the question of whether neutron stars
and their masses can be used to say anything about the existence and
properties of the H dibaryon.
In section \ref{sec:comp}, 
we discuss the condition for the occurrence of dibaryons and
strangelets in neutron star matter. The relativistic mean field model with
hyperons and the H dibaryon is presented in section \ref{sec:model}.
Implications for a H dibaryon condensate are discussed in section
\ref{sec:results} and summarized in the last section.


\section{Composite Objects in Neutron Star Matter}
\label{sec:comp}

Here we discuss the general features of the appearance of composite
quark  objects in
neutron star matter. Nuclei  will 
dissolve in dense matter due to a Mott transition at quite low density.
Hence, hypernuclei
with similar binding energies will 
also dissolve. The situation is different for 
strangelets, if they are energetically favored
compared to hadrons. Then strangelets will appear at a certain critical
density which will depend on the chemical potentials and the mass of the
strangelet. 
The most stable strangelet candidates will have a closed shell, i.e.\
they have zero total spin and are bosons. Also the H dibaryon
(consisting of two u, d, and s quarks)  has zero spin and
will form a Bose condensate if it appears in dense matter. 

The general condition for a Bose condensation of strangelets is that the
effective energy must be equal to the chemical potential:
\begin{equation}
E^*_S(k=0) = m_S + U(\rho) = B\cdot \mu_n - Z\cdot \mu_e
\end{equation}
where B stands for the baryon number and Z for the charge of the strangelet
with mass $m_S$.
The corresponding chemical potentials are $\mu_n=\mu_B$ and $\mu_e$
(unit baryon and unit negative electric
charge, respectively). 
$U(\rho)$ is the potential felt by a strangelet in a dense 
environment.
Neglecting interaction and modification of the mass in the
medium,  the threshold condition for the appearance of a strangelet
is
\begin{equation}
\frac{m_S}{B} = \mu_n - \mu_e \frac{Z}{B} \quad .
\end{equation}
Hence, the baryochemical potential determines the onset of condensation as the
charge to baryon ratio is between $+2$ and $-1$ and the electrochemical
potential is much weaker.

\begin{figure}[htb]
\begin{center}
\leavevmode
\psfig{figure=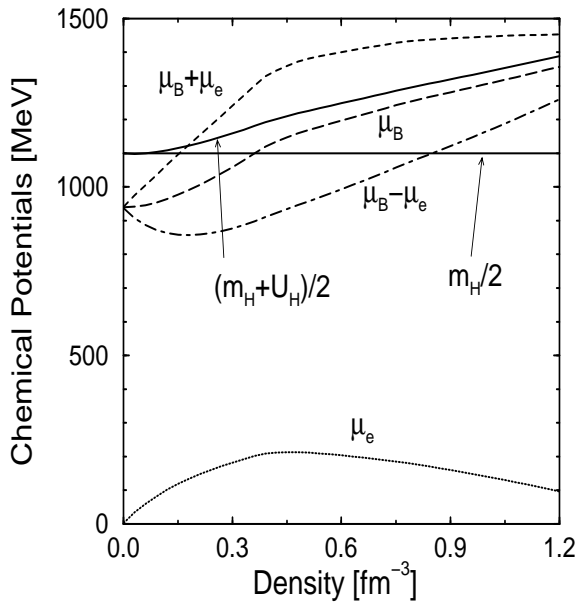,width=3.3in,height=3.3in}
\parbox[t]{4.6 in} { \caption { \label{chempot}  
The chemical potentials per baryon number
for a strangelet with a charge of $Z/A=-1$
($\mu_B + \mu_e$), 0 ($\mu_B$), and $+1$ ($\mu_B - \mu_e$).
The horizontal line is  the H dibaryon mass of $m_H = 2200$ MeV
assumed to be density independent.
The crossing of this line with the
line for $\mu_B$ 
at $\rho_c\approx 0.36$ fm$^{-3}$ marks the onset of H dibaryon
condensation. The  H mass with a medium dependent
repulsive potential of $U_H = 30$ MeV at $\rho_0$
is also shown by the curve $(m_H + U_H)/2$.
}}
\end{center}
\end{figure}

We illustrate threshold conditions for the H dibaryon (or any other
strangelet with assumed
free mass per baryon of 1.1 GeV)
in Fig.\ \ref{chempot} under two
circumstances: (1) free  dibaryon, and (2)  dibaryon whose mass is
modified by the medium. The two solid curves represent the dibaryon  mass per
baryon under these two circumstances. 
More generally, the chemical potentials for a strangelet with
$Z/B=-1$, 0, and $+1$ are also plotted. The point where these lines intersect
and rise above the mass per nucleon for a particular strangelet marks the
density threshold above which the strangelet in question would comprise
one of the constituents of matter. The threshold for free and 
medium modified dibaryons (or strangelets)
 of free mass per baryon of 1.1 GeV can be
read from the intersections.
Nucleons, leptons as well as hyperons
are included in the composition of matter.
The effect of hyperons is clearly evident in Fig.\ \ref{chempot}
through their saturating effect on $\mu_e$ at $\rho \approx 0.4$ fm$^{-3}$. 
The other chemical potentials accordingly increase
less rapidly with density above the hyperon thresholds.

The lowest mass
strangelet is the H dibaryon which is bound due to color magnetic
forces \cite{Jaffe77}.
Assuming the H has a mass of
$m_H=2.2$ GeV and does not change in the medium, gives a critical density of
about $2\rho_0$. 
Note that a negatively charged
candidate appears at a much lower density and for the assumed mass, would be
a constituent of matter at densities above $\rho \approx 0.16 {~\rm fm}^{-3}$.
If the H dibaryon feels a repulsive potential at saturation density of 
$U_H(\rho_0) =30 $ MeV
then the critical density is shifted 
beyond the maximum density
reached in the interior of a neutron star for that equation of state.
Note that for a slightly
smaller repulsive potential or for matter without hyperons 
the H dibaryon will appear (because its
chemical potential $\mu_B$ will rise above its  mass as modified by
the medium).

A similar analysis can be done for other strangelet candidates.
However, strangelets with mass numbers of $B\leq 6$ are not stable due to a
repulsive color magnetic interaction except for the H dibaryon \cite{Aerts78}.
Nevertheless, strange dibaryon states have been predicted to be bound in a
relativistic quark potential model \cite{Gold98}. Negatively charged candidates
are e.g.\ $\Sigma^-\Sigma^-$ and $\Xi^-\Xi^-$ which are heavier than the H
dibaryon but might appear at a similar density as the H due to their negative
charge. 
It has been also proposed in the MIT bag model
that negatively charged strangelets with closed
shells are likely to be most stable against strong and weak emission of
hadrons \cite{Scha97}. The candidates with a closed shell found are e.g.\
for $B=10$, $Z=-4$, for $B=12$, $Z=-6$, and $B=16$, $Z=-10$. The masses are not
precisely known as they depend crucially on the value of the bag
parameter. To be  metastable, their 
masses  per baryon should lie between $ m_\Xi$ and $m_n$. (Absolute stability
is unlikely because of the finite size shell
effect on the quark wave functions.)
Mass estimates ranges between $m_S/B=1.04-1.24$ GeV for the
above candidates. 
>From  Fig.\ \ref{chempot},
the negatively charged strangelets would  appear in neutron stars 
at densities above $ \approx 0.4 {~\rm fm}^{-3}$
if their masses are lower than
\begin{equation}
\frac{m_S}{B} \leq \mu_B^{max} + \frac{1}{2} \mu_e^{max} \approx 
1.23 \mbox{ GeV } 
\end{equation}
(assuming no interactions).
Therefore, if strangelets do not feel too high a
repulsive potential in the
medium they can appear as a  Bose condensate in neutron stars.
In the following we will discuss the modification of the properties of neutron
stars due to the appearance of a strangelet condensate. We choose to study the
case of the H dibaryon as it is the lightest candidate and might appear
first in dense matter.


\section{Mean--Field Model with Dibaryons}
\label{sec:model}

First note that the H dibaryon is a boson with zero spin and isospin.
We use the standard extended $\sigma-\omega-\rho$  model
to describe the baryon sector interacting through the mesons
\begin{eqnarray}
{\cal L} &=&
  \sum_B \bar \Psi_B \left(i \gamma_\mu \partial^\mu - m_B + g_{\sigma B}
  \sigma - g_{\omega B} 
  \gamma_\mu V_\mu - g_{\rho B} \vec{\tau}_B \vec{R}_\mu \right) \Psi_B
+ \frac{1}{2} \partial_\mu \sigma \partial^\mu \sigma - \frac{1}{2}
m_\sigma^2 \sigma^2 \cr 
 && - U(\sigma) 
- \frac{1}{4} V_{\mu\nu} V^{\mu\nu} + \frac{1}{2}m_\omega^2 V_\mu V^\mu 
+ U(V) 
- \frac{1}{4} \vec{R}_{\mu\nu} \vec{R}^{\mu\nu} 
+ \frac{1}{2}m_\rho^2 \vec{R}_\mu \vec{R}^\mu 
\end{eqnarray}
where $B$ is summed over all states of  the baryon octet, the
scalar meson is denoted by $\sigma$,
the vector mesons are denoted by
$V_\mu$ and $R_\mu$  for the iso-scalar and iso-vector meson and
$V_{\mu\nu}=\partial_\mu V_\nu -\partial_\nu V_\mu$. 
We have taken into account  possible 
self-interaction terms for the scalar field
$U(\sigma)$ \cite{Bog77}
and for the vector field \cite{Bodmer91}
\begin{equation}
U(\sigma) = \frac{1}{3} bm (g_\sigma \sigma)^3
          + \frac{1}{4} c (g_\sigma \sigma)^4\,, \qquad
U(V) = \frac{d}{4} (V_\mu V^\mu)^2\,.
\label{eq:selfint}
\end{equation}
The H dibaryon is coupled to
the mean fields by a minimal coupling scheme following \cite{Fae97a}:
\begin{equation}
{\cal L}_D = {\cal D}_\mu^* H^* {\cal D}^\mu H - {m^*_H}^2 H^*H
\label{eq:lagH}
\end{equation}
where the vector fields are coupled via the standard replacement
\begin{equation}
{\cal D}_\mu = \partial_\mu + i g_{\omega H} V_\mu \quad .
\end{equation}
This ensures consistency with Ward identities, i.e.\ the vector fields
are coupled to a conserved current. The effective mass of the H is defined as
in the baryon case 
\begin{equation}
m^*_H = m_H - g_{\sigma H} \sigma \quad .
\end{equation}
This gives, as for the vector field, a quadratic coupling term 
of the H
to the scalar
field in the Lagrangian (\ref{eq:lagH}). It turns out, that with this choice
of coupling, the scalar and vector density for the H are the same in the
mean field approximation. It was also shown
that this model is thermodynamically consistent
\cite{Fae97a}. 
The equation of motion is simply 
\begin{equation}
\left[{\cal D}^{*}_{\mu} {\cal D}^\mu 
 + {m^*_H}^2 \right] H(\omega,\vec{k}) = 0 \quad .
\end{equation}
For s-wave condensation ($\vec{k}=0$) one gets the dispersion relation
\begin{equation}
\omega_H = m^*_H + g_{\omega H} V_0 = \mu_H = 2\mu_B
\label{eq:disp}
\end{equation}
in the mean field approximation, where $\mu_B$ is the baryochemical potential.
This relation fixes the amplitude of the H dibaryon condensate. The density 
of the H dibaryon is increased until the effective energy of the H dibaryon is
equal to its chemical potential. Note that this implies that there must be 
 a 
repulsive potential between the H dibaryons at a certain H dibaryon density.
Otherwise, the effective energy is decreasing with increasing H dibaryon
density and it will never attain its chemical potential.

We do not repeat the full set of equations for the baryons
as they can be
found in e.g.\ \cite{Glen85} in detail. 
We note the additional terms due to the H dibaryon condensate in the
equations of motion
\begin{eqnarray}
m_\sigma^2 \sigma + \frac{\partial}{\partial\sigma} U(\sigma) &=&
\sum_B g_{\sigma B} \rho_s^B + 2 g_{\sigma H}  m^*_H H^* H \cr
m_\omega^2 V_0 + d V_0^3 &=&
\sum_B g_{\omega B} \rho_V^B + 2 g_{\omega H} (\mu_H - g_{\omega H} V_0) H^* H 
\quad .
\end{eqnarray}
Here one needs to define only one density for the H dibaryon 
due to the dispersion relation (\ref{eq:disp}) 
\begin{equation}
\rho_H = 2 m^*_H H^* H = 2 (\mu_H - g_{\omega H} V_0) H^* H \quad .
\end{equation}
The H dibaryon contributes to the energy density in the form
\begin{equation}
\epsilon_H = 2 {m^*_H}^2 H^*H = m^*_H \rho_H
\end{equation}
but contributes to the pressure only indirectly through the modification
of the meson fields via the
additional terms in the equations of motion.

\subsection{Baryon-Meson interactions}\label{bminter}

There exists various parameterizations in the literature for the nucleon-nucleon
interactions in the mean field model. 
The parameters are either fixed by nuclear matter properties
or fitted to properties of spherical nuclei. 
For example, the parameter set used in \cite{GM91} 
with scalar self-interactions $U(\sigma)$ 
corresponds to the nuclear matter properties:
$B/A=16.3$ MeV, $\rho_0=1.53$ fm$^{-3}$, $a_{{\rm sym}}= 32.5$
MeV, $K=240$ MeV, $m^*/m=0.78$  (which for brevity we refer to as GM91).
The parameter set TM1 \cite{Suga94}
has been fitted to the binding energy, radii, and surface thickness of 
heavy nuclei. The latter model has a
self-interaction term for the vector field $U(V)$.
This set has been used in \cite{SM96}.
We adopt these two models  as guidelines in the following. 

The hyperon
coupling constants have also been chosen differently.
We will consider two cases: (1) universal coupling of the hyperons and
(2) coupling constants using SU(6) relations.
In the former case, all hyperons are coupled equally \cite{GM91}
\begin{equation}
\frac{g_{\sigma Y}}{g_{\sigma}} = 
\frac{g_{\rho \Sigma}}{g_{\rho}} = \frac{g_{\rho \Xi}}{g_{\rho}} = 0.6
\end{equation}
where $Y$ stands for the hyperons $\Lambda$, $\Sigma$, and $\Xi$.
Note that the $\Lambda$ has isospin zero, the $\Sigma$ has isospin 1, 
while the $\Xi$ and nucleon have isospin
1/2. This gives an additional factor of 2 for the $\rho-\Sigma$ term 
in the Lagrangian and a vanishing iso-vector coupling constant for the
$\Lambda$.  

In the other case, SU(6) relations \cite{Dover84}
are used for the vector coupling constants
of the hyperons
\begin{equation}
g_{\omega} : g_{\omega \Lambda} : g_{\omega \Sigma} : g_{\omega \Xi} =
3 : 2 : 2 : 1 
\end{equation}
which scale according to the number of light quarks of the baryon.
The iso-vector coupling constants scale with the isospin like in the universal
case but are fixed differently to the nucleon iso-vector coupling constant.
In our notation this means
\begin{equation}
g_{\rho} = g_{\rho \Sigma} = g_{\rho \Xi} \,, \qquad 
g_{\rho \Lambda} = 0 \quad .
\end{equation}
The SU(6) symmetry takes already care of the isospin
so that the notation as used in \cite{SM96}
\begin{equation}
g_{\rho} : g_{\rho \Lambda} : g_{\rho \Sigma} : g_{\rho \Xi} =
1 : 0 : 2 : 1
\end{equation}
means the same. It 
reflects the strength of the iso-vector potentials of the baryons but starts
then from a different, SU(3)-symmetric Lagrangian.

Both cases, universal and SU(6), are
consistent with $\Lambda$ hypernuclear data insofar 
as the potential depth of the
$\Lambda$ in normal nuclear matter is fixed to its phenomenological value of
$U_\Lambda(\rho_0) = -30$ MeV. 
For the universal case the vector coupling constants are adjusted to this
potential depth for all hyperons. 
For the SU(6) case the scalar coupling constants are
adjusted.  Note that in addition, the SU(6) coupling scheme (quark model) 
is successful in describing the
small $\Lambda$-hypernuclear spin-orbit splitting \cite{Chia91}.

\subsection{Dibaryon interactions}

The value of the coupling constants of the H to the scalar and the 
vector field,
$g_{\sigma H}$ and $g_{\omega H}$ are unknown.
They must satisfy two constraints:
(I) the H should not appear in normal nuclear matter, and (II) the
interaction should allow for  neutron star masses at least as large as the
well established mass of the Hulse-Taylor pulsar.

The simple quark
counting rule 
suggests that $g_{\omega H}/g_{\omega N}=4/3$ as the H has four light quarks.
This choice is motivated from the success of using 
SU(6) (quark model) relations 
for hyperons in describing hypernuclear properties \cite{Dover84,Chia91}.
We use it to fix  $g_{\omega H}$.
Having fixed the vector coupling without regard to the above constraints
(I) and (II),
the burden of satisfying the constraints
falls on the scalar coupling (against
the background of the other couplings defining the Lagrangian of the
theory).

To fix the scalar coupling $g_{\sigma H}$ we consider the following.
A range for  the scalar
coupling constant $g_{\sigma H}$ can be determined from values of 
the H potential in the medium at $\rho_0$
\begin{equation}
U_H = - g_{\sigma H} \sigma + g_{\omega H} V_0
\label{U}
\end{equation}
because the meson fields are known.
The H potential at $\rho_0$
should not be deeper than
\begin{equation}
U_H(\rho_0)
 > 2 E_F - m_H = 2(m_N - 16 \mbox{ MeV}) - m_H \approx -350 \mbox{ MeV}
\end{equation}
else the  H dibaryon would condense at normal nuclear matter density.
We choose specific  discrete values in the range of
$U_H(\rho_0) = +30$, 0, --30 MeV. To each, a specific value of 
$g_{\sigma H}$ is implied through Eq.\ (\ref{U}). We will find that
potentials deeper than $-30$ MeV would decrease
the limiting neutron star mass below observed masses (at least in the
parameterizations of the Lagrangian that we have considered).

\begin{figure}[htb]
\begin{center}
\leavevmode
\psfig{figure=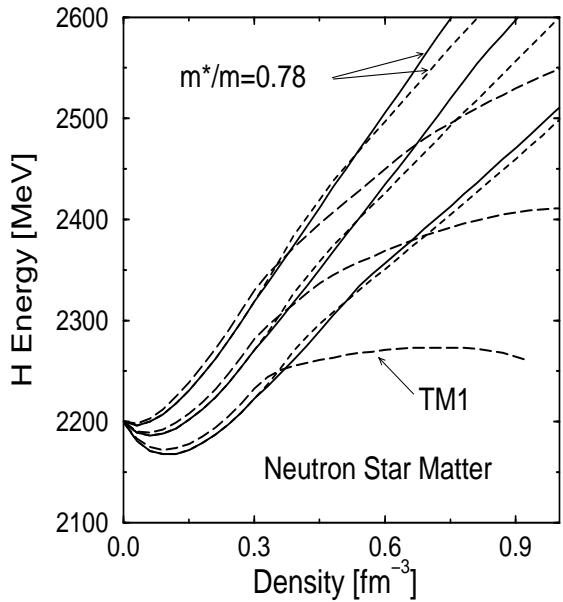,width=3.3in,height=3.3in}
\parbox[t]{4.6 in} {\caption {\label{fig:henrg} 
The energy of the H dibaryon in neutron star
matter as a function of baryon density with 
$U_H(\rho_0)=-30,0,+30$ MeV from bottom to top.
Solid lines stand for the parameter set GM91 using universal couplings, 
dotted for SU(6) couplings, dashed lines for the set TM1.}}
\end{center}
\end{figure}

The energy of the H dibaryon is plotted in Fig.\ \ref{fig:henrg}
for neutron star matter including hyperons for the various H potentials.  
The H dibaryon feels 
a repulsive potential above normal nuclear density irrespective of the chosen
potential at $\rho_0$. 
The repulsive high density behavior arises from the interaction
of the H and the vector meson. This repulsion generally dominates at
high density unless the scalar interaction is too strong. We discuss
this issue
below.
The slope at high density is quite similar for the set
GM91 and the curves for the different H potentials are just shifted. 
The vector potential dominates at high density and is chosen to be the
same (4/3 that of the nucleon) in all three cases giving the same slope at high
density. The shift comes from the 
differences in the scalar potential which saturates at high density.
The behavior at high density is quite different for the parameter set
TM1. Here the vector potential has a nonlinear dependence on the baryon density
due to the vector field selfinteraction terms. 
Note that the H energy and thus the baryon
chemical potential stays rather constant with density for the lowest curve
($U_H(\rho_0)=-30$ MeV). 
It is especially clear from Fig.\ \ref{fig:henrg} that the vacuum mass
of the H
is not as crucial to the appearance of the H in neutron stars as its
interactions with vector and scalar mesons.

In place of the above considerations for fixing a range for the scalar
coupling constant,
Faessler et al.\ \cite{Fae97a} invoked the condition 
\begin{equation}
\frac{g_{\sigma H}^2}{m_\sigma^2} < \frac{g_{\omega H}^2}{m_\omega^2}
\label{ineq}
\end{equation}
on the grounds that  otherwise the Yukawa potential between H 
dibaryons would yield a negative compressibility. This is true
at low density. 
For example, for the parameter set GM91
we find
that H matter is unstable against compression
{\sl at low density} when $U_H(\rho_0) < +2$ MeV. For the
parameter set TM1 the low density
instability arises when $U_H(\rho_0) < -10$ MeV. However,
the situation is more complicated. Even if the
equation of state has a negative slope at low density, it can become
positive at higher density. Whether or not, 
cannot be stated in terms of the inequality
of Eq.\ (\ref{ineq}) but involves all of the other interactions
and particle types in matter. Test of the low density behavior is therefore
insufficient. 

\begin{figure}[htb]
\begin{center}
\leavevmode
\psfig{figure=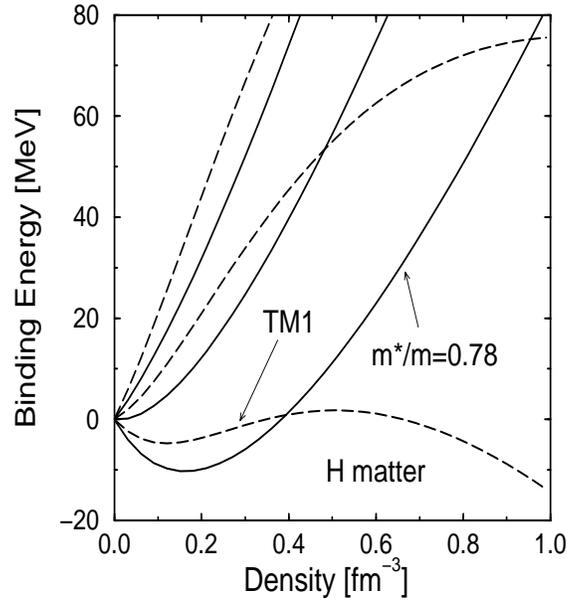,width=3.3in,height=3.3in}
\parbox[t]{4.6 in} { \caption { \label{hmatterall} The equation of state 
of pure H matter for various potential depths in
nuclear matter ($U_H(\rho_0)=-30,0,+30$ MeV from bottom to top).
Solid lines stand for the parameter set GM91, dashed lines for
TM1.
}}
\end{center}
\end{figure}

Because  gravity compresses a star, the question of stability arises
for condensed matter, not for low density matter.
For the above quoted models
the scalar and vector potentials have a different (nonlinear) 
behavior at high density which can alter the low-density conclusion. 
This is
demonstrated in Figure~\ref{hmatterall}. The binding energy of pure H
dibaryon matter is shown for the two parameter sets and the three different
choices of the H dibaryon potential in nuclear matter.
One sees that the compressibility, which is proportional to the slope of the
curves, is negative at low density for $U_H(\rho_0)=-30$ 
MeV for both parameter sets.
Nevertheless, the scalar self-interactions provide a nonlinear dependence of the
scalar potential on the density (here H dibaryon number density). This results
in an overall repulsive potential for pure H matter at higher density and a
minimum around normal nuclear density. Hence H  matter can be stable
at high density even if the low density limit seems to indicate an instability.

With reference to Fig.\ \ref{hmatterall}, we see that even more
complicated situations can arise.
For the parameter set TM1, the vector self-interactions cause 
the energy density due to the  vector potential
to rise like $\rho^{1/3}$  instead of like $\rho$, as for the linear
behavior in the standard Walecka model. 
It can then happen that  the scalar potential wins ultimately
over the vector potential at very high density.
Such a case is seen in Figure~\ref{hmatterall} for the lowest
dashed curve: the equation of state has a local minimum around normal nuclear
density but the compressibility becomes
negative again at higher density.

The instability for pure H matter does not mean that a neutron star with a H
condensate is unstable against compressional modes, as the overall
compressibility can still be positive. This will depend on the 
intrinsic stiffness of the
equation of state
 used and the hadron population inside the neutron star. 
A recent analysis considered neutron star matter
consisting of nucleons and leptons only \cite{Fae97b} and found rather
stringent conditions for the coupling constant of the H dibaryon.
The appearance of hyperons already at $(2-3)\rho_0$ \cite{Glen85} will
certainly alter their conclusions and will be discussed in the following.


\section{H Dibaryons in Neutron Stars}
\label{sec:results}

\subsection{Populations}
To give an early impression of the possible presence of H dibaryons
in neutron stars under acceptable conditions as to (1) its absence
in normal matter, and (2) an acceptable value for the limiting mass
neutron star, we compare the populations of the limiting mass star in the
absence of the dibaryons and in their presence, in Figs.\ 
\ref{comp_h_k240m78_su6} and \ref{comp_h_k240m78_su6_uhm30}.
\begin{figure}[tbh]
\vspace{-.5in}
\begin{center}
\leavevmode
\centerline{ \hbox{
\psfig{figure=ps.comp_h_k240m78_su6,width=3.in,height=3.6in}
\hspace{.1in}
\psfig{figure=ps.comp_h_k240m78_su6_uhm30,width=3in,height=3.6in}
}}
\begin{flushright}
\parbox[t]{2.7in} { \caption { \label{comp_h_k240m78_su6} Populations of
octet baryons and leptons in a limiting mass star with nuclear properties
as described by the case GM91 with hyperon couplings chosen as in the
SU(6) scheme.
}} \ \hspace{.4in} \
\parbox[t]{2.7in} { \caption { \label{comp_h_k240m78_su6_uhm30} Similar 
to Fig.\  \protect\ref{comp_h_k240m78_su6} but with the H dibaryon
experiencing a potential of $U_H(\rho_0)=-30$ MeV.
}}
\end{flushright}
\end{center}
\end{figure}
Hyperons appear abundantly in the interior
of the neutron star. The hyperons $\Lambda$ and $\Xi^-$ reach values close to
the proton density in the stellar  core. Protons are more abundant
once the negatively charged hyperons $\Sigma^-$ and $\Xi^-$ are present to
compensate the charge of the proton.
As can be seen in Fig.\
\ref{comp_h_k240m78_su6_uhm30}, 
particle populations are strongly modified in the core of the star where the 
dibaryon appears. The proton population is suppressed
since baryon number is carried more in the H bosons.
Likewise the hyperon populations are strongly suppressed. This is not to say
that these baryons have little influence on the H presence. It is in
matter containing significant $\Sigma^-$ and $\Lambda$ populations that
the H threshold is attained.
Outside the core, beyond $r\approx 4$ km, populations
and the stellar radius are hardly changed by the presence of
the H in the core and therefore 
cooling of the star
would  be little effected by the H.
H dibaryons could lie within the
star while providing no  direct sign of their presence.

\subsection{Limits on the H dibaryon mass}

\begin{figure}[htb]
\begin{center}
\leavevmode
\psfig{figure=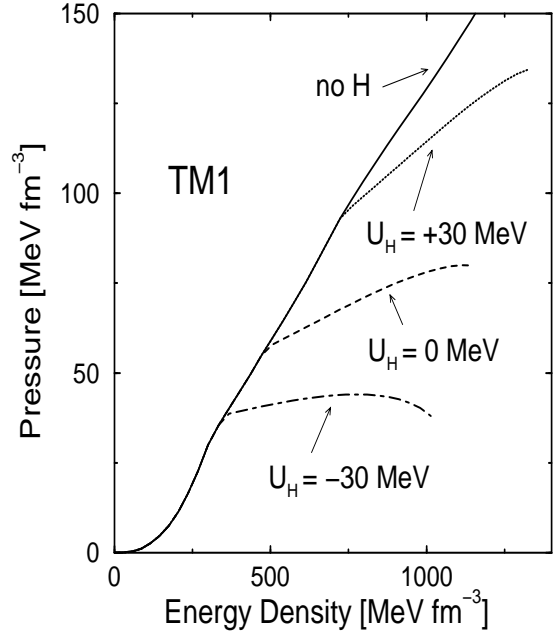,width=3.3in,height=3.3in}
\parbox[t]{4.6 in} { \caption { \label{eos_tm1} Equation of state
for the parameter set TM1, SU(6) coupling for hyperons 
 and several values of the H dibaryon potential $U_H(\rho_0)$.
}}
\end{center}
\end{figure}

Generally the maximum mass of a neutron star is lowered 
due to H dibaryon condensation
because
the condensate does not contribute directly to the pressure and it
removes the pressure due to two baryons per dibaryon in the condensate.
The
equation of state is thereby softened.
As the maximum mass should be not lower than $1.44M_\odot$ (the mass of the
Hulse-Taylor pulsar),
one can try to impose constraints on the {\sl medium} mass of the H
dibaryon. 
For the nuclear parameterization TM1, we show the equation of state
corresponding to several values of $U_H (\rho_0)$ in Fig.\ \ref{eos_tm1}.
The equation of state can be considerably softened by the condensation of H
dibaryons. Especially for the case $U_H (\rho_0) = -30$ MeV the pressure stays
nearly constant once condensation starts. The H boson constitutes a large
fraction of the matter but does not contribute to the pressure directly. The
plateau seen in  the equation of state in Fig.\ \ref{eos_tm1} can be traced
back to the equation of state for pure H dibaryon matter in 
Fig.~\ref{hmatterall} (see the bottom dashed curve). 

Figure \ref{mas_hdibaryon} summarizes the neutron star masses for
the two hyperon coupling schemes SU(6) and `universal' (see section
\ref{bminter}) and for
various values of the interaction
$U_H(\rho_0)$.  The more attractive the interaction, the more populous the
dibaryon, the softer the equation of state
and the smaller the limiting mass.
The potential $U_H(\rho_0)=-30$ MeV 
is about as attractive as is compatible with 
the Hulse-Taylor pulsar. For the SU(6) coupling, H dibaryons feeling
$U_H(\rho_0) = 0$ MeV would not
be present in the stable members of the sequence, while for universal 
coupling of hyperons, the dibaryon is present in small number and reduces the
limiting mass marginally by $\approx 0.03 M_\odot$. A repulsive dibaryon
interaction in the medium would therefore assure its minimal presence
if not its total absence (as  is the case for the SU(6) coupling).

\begin{figure}[tbh]
\vspace{-.5in}
\begin{center}
\leavevmode
\centerline{ \hbox{
\psfig{figure=ps.mas_hdibaryon,width=3.in,height=3.6in}
\hspace{.1in}
\psfig{figure=ps.mass_hd_d,width=3in,height=3.6in}
}}
\begin{flushright}
\parbox[t]{2.7in} { \caption { \label{mas_hdibaryon} Details of neutron
star sequences near the limiting mass for the nuclear parameterization 
GM91 and two hyperon coupling schemes
labeled `universal' and SU(6) and for each of these, several
values of the dibaryon interaction $U_H (\rho_0)$.
}} \ \hspace{.4in} \
\parbox[t]{2.7in} { \caption { \label{mass_hd_d} Similar to
Fig.\ \protect\ref{mas_hdibaryon} but for the nuclear coupling 
denoted by TM1.
}}
\end{flushright}
\end{center}
\end{figure}

One might infer from the figure, that very
attractive potentials for the H of $U_H(\rho_0)< -30$ MeV 
can be ruled out by neutron
star data. Nevertheless, such a conclusion must be moderated by
our ignorance of the H dibaryon mass.
The appearance of the H condensation depends on the in-medium 
potential at high density and the mass of the H
dibaryon which are unknown.

In addition to uncertainties in the vacuum and medium mass of the 
H, there are uncertainties in the underlying nuclear equation of state
especially 
at densities above normal nuclear density. 
This can be seen by contrasting 
Fig. \ref{mas_hdibaryon} with Fig.\ \ref{mass_hd_d}. In the second
case the nuclear parameterization is TM1 of section \ref{bminter}.
Here we see that the H is present provided $U_H(\rho_0)<+30$ MeV.
For an attractive potential of
$U_H(\rho_0)=-30$ MeV, 
one finds that the mass of the neutron star reaches a plateau,
i.e.\ the mass of the neutron star is independent of the central energy
density. The maximum mass is then $1.44M_\odot$.
This is the lowest value allowed by present observation.
The plateau comes from the fact, that the equation of
state has a nearly constant pressure for the
particular coupling due to the appearance of the H dibaryon.
A more attractive potential than $U_H(\rho_0)=-30$ MeV
would result in a negative curvature of the
pressure and therefore in an unstable equation of state. Note, that this
behavior is
mainly related to the potential between the H dibaryons
as discussed in connection with pure H dibaryon matter
(see Figure~\ref{hmatterall}).
Neutron star matter is stabilized against collapse to H dibaryon
condensate because of the repulsive interaction with the
vector meson provided the H dibaryon density is not too large.
The presence of the other
baryons in the matter tends to stabilize the system compared to pure H matter.

\begin{figure}[tbh]
\vspace{-.5in}
\begin{center}
\leavevmode
\psfig{figure=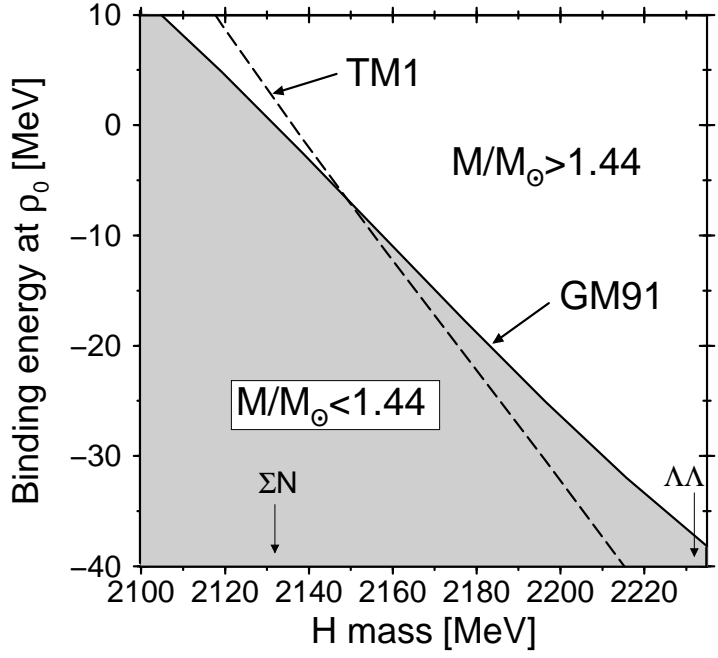,height=3.6in}
\parbox[t]{4.6 in}{\caption { \label{fig:hcrit2}
Diagram of the H vacuum mass and the H binding energy in
normal nuclear matter. The shaded region gives a maximum neutron star mass
lower than $1.44 M_\odot$ for the parameter set GM91 using universal coupling,
the dotted line denotes the case for TM1.  
}}
\end{center}
\end{figure}

In the present approach one can now exclude certain regions of the assumed mass
of the H Dibaryon $m_H$ and its potential at saturation density $U_H$. 
For a too low H mass
or a too deep potential the presence of the H dibaryon condensate will lower
the maximum mass of a neutron star below the observed limit of $1.44M_\odot$.
This excluded region is shown as a shaded area 
in Fig.\ \ref{fig:hcrit2} in an $(U_H, m_H)$ diagram for the parameter set
GM91 using universal coupling. 
The dashed line denotes the case for the parameter set TM1.
The thresholds for $\Lambda\Lambda$ and $\Sigma$N decay which are relevant for
the lifetime of the H dibaryon \cite{Don86} are also indicated.
The results for the two parameterizations are quite close
to each other despite their different high density
behavior. For an H mass of lower than 2.13 GeV 
(the $\Sigma$N threshold)
an attractive potential in
nuclear matter gives too low maximum neutron star mass. 
Hence, rather long-lived H dibaryons ($\tau> 10^{-7}$ s according to
\cite{Don86}) are unlikely to form bound H hypernuclear states.
On the other hand, attractive potentials lower than $U_H(\rho_0)<-50$ MeV seem
to be ruled out by neutron star mass constraints for the H mass range
considered. Otherwise, the H dibaryon has 
to be heavier than the 2$\Lambda$ threshold and will be a resonance state.
These limits will depend also on the chosen vector
coupling constant. We are using an effective model and  
the extrapolation to 
high density might be completely different in reality.

\subsection{Radius and the Mass-Radius Relation}

The mass-radius relation for the GM91 model with universal 
hyperon couplings is shown in Fig.\ \ref{rm_hu_3} for several
values of the dibaryon potential $U_H(\rho_0)$. 
The more attractive the potential,
the softer the equation of state
and the lower the limiting mass, as remarked earlier. 
The radius of the limiting star decreases the more attractive the potential
$U_H(\rho_0)$ because the star, having less mass,
is less gravitationally compacted.
Similar results are shown in Fig.\ \ref{rm_hu_full} for the
nuclear parameterization TM1.
In this case the limiting mass stars have substantially larger radii.
\begin{figure}[tbh]
\vspace{-.5in}
\begin{center}
\leavevmode
\centerline{ \hbox{
\psfig{figure=ps.rm_hu_3,width=3.in,height=3.6in}
\hspace{.1in}
\psfig{figure=ps.rm_hu_full,width=3in,height=3.6in}
}}
\begin{flushright}
\parbox[t]{2.7in} { \caption { \label{rm_hu_3} Mass-radius relation
for the nuclear model GM91, universal hyperon coupling
 and several values of the dibaryon
potential $U_H(\rho_0)$.
}} \ \hspace{.4in} \
\parbox[t]{2.7in} { \caption { \label{rm_hu_full} Similar
to Fig.\  \protect\ref{rm_hu_3} but for the nuclear model TM1.
}}
\end{flushright}
\end{center}
\end{figure}

It is interesting to note, that the presence of the H dibaryon in neutron stars
seems to lower the maximum mass but increases the minimum radius. The
mass--radius relation just stops at the point where the H dibaryon condensation
sets in. This is contrary to kaon condensation \cite{Thorsson94,Fuji96} where
the radius decreases for a kaon condensed star. Note that kaon condensation
produces a strong phase transition of first order but the equation of state has
no plateau if treated in a
thermodynamically consistent way \cite{SG98}. In addition,
baryons are not replaced by a baryon number carrying condensate but neutrons
are replaced by protons and $K^-$'s resulting in a different mass-radius
relation.


\section{Summary}
\label{sec:summary}

We are particularly interested in the question of whether neutron stars
and their masses can be used to say anything about the existence and
properties of the H dibaryon.
We have studied the influence of the possible occurrence of an H dibaryon
condensate and strangelets in neutron stars including hyperons. 
Without in-medium modifications, it is quite likely that especially negatively
charged strangelets, if they exists, will be present 
in the dense interior of neutron stars.
 
The appearance of  H dibaryons in the stellar core depends crucially on
their mass and on the chosen potential of
the H in nuclear matter. Hyperons tend to shift the onset of the H to higher
density or to prevent H dibaryon condensation. 
If the condensation happens and if the potential of the H is attractive enough
to provide a substantial number density in the neutron star, the maximum mass
of the neutron star is reduced compared to the case without the H dibaryon.
The decrease of the maximum mass is moderate and allows for the presence of H
dibaryons in the interior of neutron stars in accord with present neutron star
mass data. If the H dibaryon feels an attractive potential in matter, it can
lead to a plateau in the mass of neutron stars, as there exist a region of
very slowly rising pressure with energy density. 

 If the limiting neutron star mass
 is about that of the Hulse-Taylor pulsar
 a  condensate of H dibaryons of vacuum mass $\sim 2.2$ GeV
 and a moderately attractive potential in the medium
 could not be ruled out.
 On the other hand, if the medium potential were even moderately repulsive,
 the H,  would not likely exist  in neutron stars.
 If neutron stars of mass $\sim 1.6 M_\odot$ were known to exist,
 attractive medium effects could be ruled out.
 For a mass limit of $1.44 M_\odot$, attractive potentials for an H mass below
 the $\Sigma$N threshold (1.3 GeV) are ruled out.  

H dibaryon or strangelet condensation 
might happen as a precursor to the phase transition to a quark
plasma. In this respect, we note that this phase transition is of first order
\cite{Glen92}. Hence, small bubbles of strange quark matter will appear in the
mixed phase which are most likely negatively charged due to the isospin
potential of the nuclear matter. This is in line with the results presented
here. 
As the most stable strangelets have spin zero, the onset to a quark plasma 
will be initiated by a Bose condensation of strangelets (possibly including the
H dibaryon). As the phase transition proceeds, the bubbles will overlap and
will finally 
replace nuclear matter by essentially filling up the whole volume.

~~\\[2ex]
Acknowledgments:

J.S.B. acknowledges support
by the Alexander-von-Humboldt Stiftung with a Feodor-Lynen fellowship.
This work is supported by 
the Director, Office of Energy Research,
Office of High Energy and Nuclear Physics, Nuclear Physics Division of the
U.S. Department of Energy under Contract No.\ DE-AC03-76SF00098.




\begin{thebibliography}{10}

\bibitem{Jaffe77}
R.~L. Jaffe, Phys. Rev. Lett. {\bf 38},  195, 617(E)  (1977).

\bibitem{Carroll78}
S.~A. Carroll {\it et~al.}, Phys. Rev. Lett. {\bf 41},  777, 1002(E)  (1978).

\bibitem{Aoki90}
S. Aoki {\it et~al.}, Phys. Rev. Lett. {\bf 65},  1729  (1990).

\bibitem{Belz96}
J. Belz {\it et~al.}, Phys. Rev. Lett. {\bf 76},  3277  (1996).

\bibitem{Belz97}
J. Belz {\it et~al.}, Phys. Rev. C {\bf 56},  1164  (1997).

\bibitem{Stotz97}
R.~W. Stotzer {\it et~al.}, Phys. Rev. Lett. {\bf 78},  3646  (1997).

\bibitem{Craw98}
{H. J. Crawford (E896 collaboration)}, Nucl. Phys. A  in press  (1998).

\bibitem{Shahba93}
B.~A. Shahbazian, T.~A. Volokhovskaya, V.~N. Emelyanenko, and A.~S. Martynov,
  Phys. Lett. B {\bf 316},  593  (1993).

\bibitem{Long95}
R.~S. Longacre {\it et~al.}, Nucl. Phys. A {\bf 590},  477c  (1995).

\bibitem{HYP97}
{Proceedings of the International Conference on Hypernuclear and Strange
  Particle Physics (HYP'97)}, Nucl. Phys. A  to be published  (1998).

\bibitem{Dover89}
C.~B. Dover, Nuovo Cim. {\bf 102A},  521  (1989).

\bibitem{Dalitz89}
R.~H. Dalitz {\it et~al.}, Proc. Roy. Soc. Lond. {\bf A426},  1  (1989).

\bibitem{Danysz63}
M. Danysz {\it et~al.}, Nucl. Phys. {\bf 49},  121  (1963).

\bibitem{Prowse66}
D.~J. Prowse, Phys. Rev. Lett. {\bf 17},  782  (1966).

\bibitem{Aoki91}
S. Aoki {\it et~al.}, Prog. Theor. Phys. {\bf 85},  1287  (1991).

\bibitem{Dover91}
C.~B. Dover, D.~J. Millener, A. Gal, and D.~H. Davis, Phys. Rev. C {\bf 44},
  1905  (1991).

\bibitem{Kerb84}
B.~O. Kerbikov, Sov. J. Nucl. Phys. {\bf 39},  516  (1984).

\bibitem{Jaffe91}
R.~L. Jaffe, Nucl. Phys. B (Proc. Suppl.) {\bf 24B},  8  (1991).

\bibitem{Glen85}
N.~K. Glendenning, Astrophys. J. {\bf 293},  470  (1985).

\bibitem{Fae97a}
A. Faessler, A.~J. Buchmann, and M.~I. Krivoruchenko, Phys. Lett. B {\bf 391},
  255  (1997).

\bibitem{Fae97b}
A. Faessler, A.~J. Buchmann, and M.~I. Krivoruchenko, Phys. Rev. C {\bf 56},
  1576  (1997).

\bibitem{Arm97}
T.~A. Armstrong {\it et~al.}, Phys. Rev. Lett. {\bf 79},  3612  (1997).

\bibitem{Beavies95}
D. Beavies {\it et~al.}, Phys. Rev. Lett. {\bf 75},  3078  (1995).

\bibitem{Apple96}
G. Applequist {\it et~al.}, Phys. Rev. Lett. {\bf 76},  3907  (1996).

\bibitem{Aerts78}
A.~T.~M. Aerts, P.~J.~G. Mulders, and J.~J. de~Swart, Phys. Rev. D {\bf 17},
  260  (1978).

\bibitem{Gilson93}
E.~P. Gilson and R.~L. Jaffe, Phys. Rev. Lett. {\bf 71},  332  (1993).

\bibitem{Scha97}
J. Schaffner-Bielich, A. Diener, C. Greiner, and H. St\"ocker, Phys. Rev. C
  {\bf 55},  3038  (1997).

\bibitem{GS96}
C. Greiner and J. Schaffner, Int. J. Mod. Phys. E {\bf 5},  239  (1996).

\bibitem{Gold98}
T. Goldman {\it et~al.}, Systematic Theoretical Search for Dibaryons in a
  Relativistic Model, nucl-th/9803002, 1998.

\bibitem{Bog77}
J. Boguta and A.~R. Bodmer, Nucl. Phys. A {\bf 292},  413  (1977).

\bibitem{Bodmer91}
A.~R. Bodmer, Nucl. Phys. A {\bf 526},  703  (1991).

\bibitem{GM91}
N.~K. Glendenning and S.~A. Moszkowski, Phys. Rev. Lett. {\bf 67},  2414
  (1991).

\bibitem{Suga94}
Y. Sugahara and H. Toki, Nucl. Phys. A {\bf 579},  557  (1994).

\bibitem{SM96}
J. Schaffner and I.~N. Mishustin, Phys. Rev. C {\bf 53},  1416  (1996).

\bibitem{Dover84}
C.~B. Dover and A. Gal, Progr. Part. Nucl. Phys. {\bf 12},  171  (1984).

\bibitem{Chia91}
M. Chiapparini, A.~O. Gattone, and B.~K. Jennings, Nucl. Phys. A {\bf 529},
  589  (1991).

\bibitem{Don86}
J.~F. Donoghue, E. Golowich, and B.~R. Holstein, Phys. Rev. D {\bf 34},  3434
  (1986).

\bibitem{Thorsson94}
V. Thorsson, M. Prakash, and J.~M. Lattimer, Nucl. Phys. A {\bf 572},  693
  (1994).

\bibitem{Fuji96}
H. Fujii, T. Maruyama, T. Muto, and T. Tatsumi, Nucl. Phys. A {\bf 597},  645
  (1996).

\bibitem{SG98}
J. Schaffner-Bielich and N.~K. Glendenning, Strange Phases in Neutron Stars,
  Proceedings of the Workshop on Nuclear Astrophysics, Hirschegg, Austria,
  1998, nucl-th/9802030.

\bibitem{Glen92}
N.~K. Glendenning, Phys. Rev. D {\bf 46},  1274  (1992).

\end{thebibliography}
\end{document}